\documentclass[lettersize,journal,twoside]{IEEEtran}
\usepackage[utf8]{inputenc}
\usepackage{amsmath,amsfonts}
\usepackage{algorithmic}
\usepackage{algorithm}
\usepackage{array}
\usepackage[caption=false,font=footnotesize,labelfont=rm,textfont=rm]{subfig}
\usepackage{color}
\usepackage{textcomp}
\usepackage{stfloats}
\usepackage{url}
\usepackage{verbatim}
\usepackage{graphicx}
\usepackage{cite}
\usepackage{tabularx}
\usepackage{gensymb}
\def\BibTeX{{\rm B\kern-.05em{\sc i\kern-.025em b}\kern-.08em
    T\kern-.1667em\lower.7ex\hbox{E}\kern-.125emX}}

\begin{document} 

\title{Direct-to-Cell Satellite Communications: Task-Oriented Paradigm for 6G Closed-Loop Autonomous Operations}

\author{Daohong Shen, Wei Feng,~\IEEEmembership{Senior Member,~IEEE}, Yunfei Chen,~\IEEEmembership{Fellow,~IEEE}, Yongxu Zhu,~\IEEEmembership{Senior Member,~IEEE}, Jinxia Cheng, Dapeng Wang, Shi Jin,~\IEEEmembership{Fellow,~IEEE}

\thanks{Daohong Shen and Wei Feng (corresponding author) are with the Department of Electronic Engineering, State Key Laboratory of Space Network and Communications, Tsinghua University, Beijing 100084, China. Yunfei Chen is with the Department of Engineering, University of Durham, DH1 3LE Durham, U.K. Yongxu Zhu and Shi Jin are with the National Mobile Communications Research Laboratory, Southeast University, Nanjing 210096, China.
Yongxu Zhu is also with the Purple Mountain Laboratories, Nanjing 211111, China.
Jinxia Cheng and Dapeng Wang are with the China Mobile Research Institute, Beijing 100053, China.}}

\maketitle

\begin{abstract}

Direct-to-cell (D2C) satellite communications are emerging as a complement to terrestrial communications in the sixth generation (6G) mobile networks due to their wide-area coverage capability. 
In addition to human-oriented D2C services for handheld devices, future autonomous operations may involve robotic terminals, such as robotic platforms and vehicular devices, to directly access satellite networks for task continuity.
Human-oriented communications mainly focus on link-level quality of service. 
In contrast, robotic operations require task-oriented designs, where reliable task completion is the ultimate objective. 
Therefore, sensing, communication, computing, and control should be evaluated jointly rather than separately in such applications.
Motivated by this, we model the system with a sensing–communication–computing–control (SC$^3$) closed loop and analyze it from an entropy perspective. 
Furthermore, to manage the complexity of the closed-loop network, we decompose it into fine-grained functional structures and investigate the key challenges of collaborative sensing, collaborative computing, and collaborative control. 
A case study is presented to illustrate the potential benefit of the proposed task-oriented design over conventional communication-oriented schemes in terms of control cost.
Finally, several open issues are outlined for future research and practical implementation.
\end{abstract}


\section{Introduction}
\begin{figure}[ht]
\centering
\includegraphics[width=3.2in]{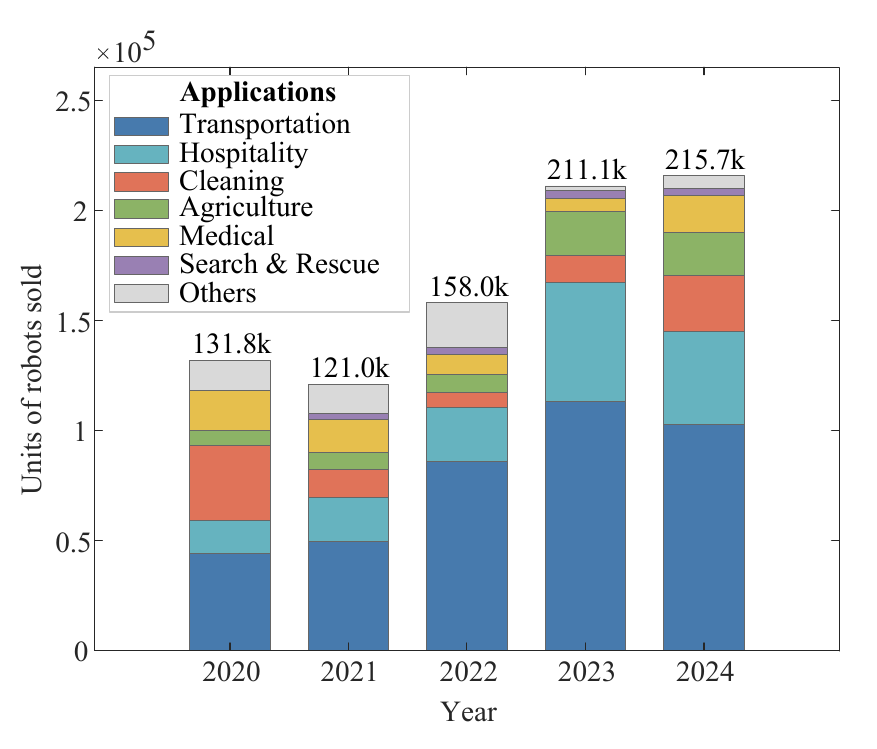}
\caption{Professional service robot sales registered by the International Federation of Robotics (IFR)  Statistical Department (2020–2024).}
\label{fig_01}
\end{figure}

With the rapid development of embodied artificial intelligence (AI) and automation technologies, autonomous operations using intelligent robots are becoming a dominant paradigm in modern industrial systems. 
As shown in Fig.~\ref{fig_01}, sales of service robots, which can be used in transportation, professional cleaning, agriculture, and search \& rescue, have grown significantly in recent years~\cite{survey}.
In cases such as emergency rescue and remote industrial applications, autonomous operations have demonstrated significant advantages in terms of safety, efficiency, and scalability, enabling effective management over large and complex areas~\cite{case}.
However, in many of these scenarios, terrestrial communication infrastructures are either unavailable or severely damaged due to harsh environments or unexpected disasters, posing a critical challenge to reliable system coordination and control.
In such cases, direct-to-cell (D2C) communications can complement terrestrial networks, providing wide-area connectivity and maintaining the continuity of closed-loop autonomous operations.
It should be noted that the term D2C usually refers to satellite connectivity to standard handheld cellular devices with low-gain antennas and limited transmit power. 
In contrast, the autonomous operation scenarios considered in this article may adopt enhanced terminals mounted on sensors, robots, vehicles, or mission-specific devices, so that D2C is used in this article in a robot-based sense.

Existing works on D2C technologies primarily focus on the interconnection between satellites and humans, where communications performance, such as data rate and reliability, is the primary metric for optimization.
However, 6G marks a pivotal shift from ``connected people" to ``connected intelligence".
As shown in Table~\ref{tab:table1}, there are many noticeable differences between task-oriented and human-oriented D2C communications.
Human-oriented D2C services, such as non-terrestrial internet access, mainly focus on user experience metrics including throughput and coverage, where data rates are typically at the Mbps level and latency requirements are relatively relaxed. 
In contrast, task-oriented autonomous operations involve mission-critical control tasks, which may require latency at the tens-of-milliseconds level and ultra-high reliability with block error rates on the order of \(10^{-5}\) to ensure control stability~\cite{requirements}. 
Therefore, in autonomous operations, the network is expected not only to deliver bits, but also to support task completion, control stability, and system-level operational performance.
Consequently, these scenarios motivate a shift from communication-oriented optimization to task-oriented design for robot-dominant D2C communications.

\IEEEpubidadjcol

\begin{table*}[!t]
    \caption{Comparison of features between task-oriented and human-oriented D2C communications.}
    \begin{tabularx}{\textwidth}{lXXX} 
    \hline
      \textbf{} & \textbf{Task-oriented D2C} & \textbf{Human-oriented D2C}\\
      \hline
      \textbf{Main application scenarios }&  Search and rescue, transportation and logistics, agriculture,  professional cleaning, inspection, etc  & In-flight Internet access, maritime communications, Internet access in remote areas or exploration, etc\\
      \textbf{Data characteristics}  &  More regularly generated, high density, long-term service, need to be processed, etc  & More randomly generated, multimedia data, high latency requirements, etc  \\
      \textbf{Key evaluation metrics}  &  Task completion, control performance, system stability, etc  &  Data rate, coverage, energy consumption, etc\\
    \hline
    \end{tabularx}
  \label{tab:table1}
\end{table*}

Autonomous operations involve a sequence of tightly coupled stages, including information acquisition, data transmission, data processing and physical control.
Typically, sensors collect environmental and target-state information, communication links deliver sensing data and control commands, computing units process the received data and generate decisions, and robots execute the commands in the physical world.
These stages form a sensing-communication-computing-control (SC$^3$) closed loop, where impairments in any stage may propagate throughout the loop, degrading overall task performance.
Furthermore, since the total resource is limited, all stages must compete in a system-level optimization.
Pei et al. integrate routing, resource scheduling, edge computing, and latency prediction to meet diverse communication task requirements in 6G non-terrestrial networks (NTNs)~\cite{task}. 
However, the relationship between these network capabilities and the performance of the controlled physical system is not explicitly characterized. 
This leaves open how sensing, communication, and computing resources should be jointly managed to support closed-loop autonomous operations.

Motivated by the above observations, we investigate task-oriented design for robotic D2C communications in 6G closed-loop autonomous operations.
We present a satellite-assisted SC$^3$ closed-loop architecture for autonomous operations, where satellite communications are positioned as a complement to terrestrial networks in remote or infrastructure-impaired areas.
An entropy-based performance evaluation is introduced to characterize the completion of tasks with inherent coupling mechanisms across different stages, based on which a system-level optimization is developed to enhance the closed-loop performance.
To gain deeper insights into the system, we make an analogy to the neural system and take the single closed loop as the basic component. 
We further consider the collaborative sensing, collaborative computing, and collaborative control structures and investigate the fundamental issues within each structure.
A case study is given to illustrate the potential benefit of task-oriented design over conventional communication-oriented schemes.
Finally, several promising directions for future research are discussed.

\section{D2C satellite communications with SC$^3$ closed loops}
\begin{figure*}[ht]
\centering
\includegraphics[width=7in]{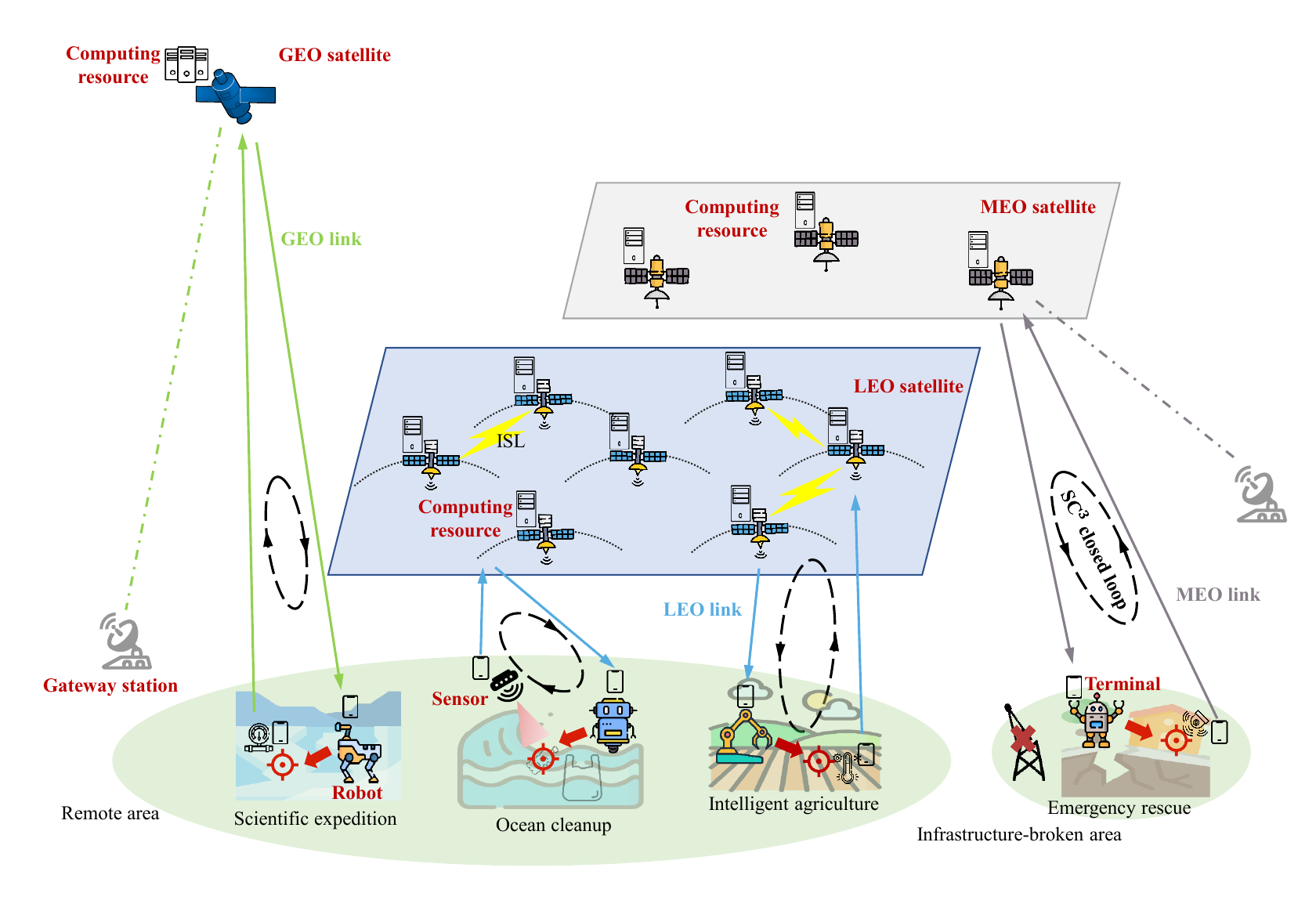}
\caption{Illustration of robot-based D2C satellite communications for closed-loop operations.}
\label{fig_1}
\end{figure*}

The overview of the robot-based D2C satellite communication system for closed-loop operations is shown in Fig.~\ref{fig_1}.
To more efficiently characterize the system, we analyze it from an entropy perspective and present the corresponding system design methodologies.

\subsection{SC$^3$ closed loop}
As shown in Fig.~\ref{fig_1}, a variety of autonomous operational tasks, such as intelligent agriculture, ocean cleanup, scientific expedition, and emergency rescue, are supported by different robots, which are sent to target areas based on their specific functions. 
In these scenarios, both sensors and robots are equipped with D2C terminals that can directly communicate with satellites.
During task execution, sensors collect environmental information and target-state data. 
The collected data are then uploaded to the satellites via D2C links. 
Geostationary Earth orbit (GEO), medium Earth orbit (MEO) and low Earth orbit (LEO) satellites can all receive data. 
There are many metrics for measuring task requirements, such as delay, energy consumption, and user privacy~\cite{requirements}. 
Different tasks typically impose distinct requirements; for example, intelligent agriculture requires low latency, while disaster relief requires a low block error rate~\cite{IOT}.
According to task requirements and available resources, the data can be processed onboard, cooperatively across satellites through inter-satellite links (ISLs), or forwarded via gateways to ground servers~\cite{ntn}. 
The resulting control commands are then delivered to robots through downlink D2C links, and the robots take appropriate actions to accomplish the assigned operational tasks. 
This tightly integrated sequence of ``sensing, communication, computing, and control" forms an SC$^3$ closed loop. 
Efficient information exchange within this closed loop directly enhances the effectiveness of autonomous operations but is not the ultimate goal, as the capability of each stage and their coupling jointly determine the overall performance of the system. 
Therefore, a holistic consideration of the individual stage capability and their interaction is essential for improving the performance of closed-loop operations. 

\subsection{Entropy-based characterization of system performance}
Based on the SC$^3$ closed loop, we characterize the overall system performance from an entropy perspective.
Entropy is a fundamental measure of a system's uncertainty.
In closed-loop autonomous operations, the physical state of the controlled object evolves with environmental disturbances, transmission errors, imperfect observations, and so on.
If no effective feedback information is provided, the uncertainty of the system state may continuously increase.
In control systems, the intrinsic entropy rate quantifies the uncertainty produced by the evolution of the system state over time.
From~\cite{entropy}, a system can reach a stable state only when its information rate exceeds its intrinsic entropy rate.
A larger intrinsic entropy rate indicates that more effective information is required to maintain closed-loop stability.

In a SC$^3$ closed loop, the information that finally contributes to control is jointly constrained by sensing, uplink transmission, computing, downlink transmission, and robot actuation.
For example, insufficient sensing quality limits the amount of useful environmental information, a weak uplink restricts the sensing data that can be uploaded, limited computing capability reduces the speed of task-relevant information extraction, and a constrained downlink prevents timely delivery of control commands.
Therefore, the effective information of the whole closed loop is determined by the bottleneck among all stages.
We characterize the system’s control performance by the amount of information per unit time that finally takes effect to maintain stability, referred to as the closed-loop neg-entropy rate (CNER)~\cite{CNER}.
CNER is not a pure communications rate.
Instead, it measures the amount of task-effective information per unit time.
Only when CNER exceeds the intrinsic entropy rate can the system be stabilized.

The CNER also provides an information-level interpretation of the control cost. 
The linear quadratic regulator (LQR) cost measures physical control performance by penalizing both state deviation and control effort. 
A lower LQR cost indicates better stability or lower control energy consumption. 
According to the rate-cost relationship in control theory, a lower bound of the LQR cost has the following relationship with the CNER~\cite{lqr}:
\begin{equation}
l \geq \frac{nN(\mathbf{v}) \left| \det \mathbf{M} \right|^{\frac{1}{n}}}{2^{\frac{2}{n}\left(\mathbf{CNER}-\log_{2}\left|\det\mathbf{A}\right|\right)}-1}+\operatorname{tr}\left(\mathbf{\Sigma}_{\mathbf{v}}\mathbf{S}\right),
\end{equation}
where $n$ is the dimension of the system state, $N(\mathbf{v})$ represents the entropy power of the process noise $\mathbf{v}$, 
$\mathbf{\Sigma}_{\mathbf{v}}$ is the covariance matrix of the process noise,
$\mathbf{A}$ is the state transition matrix,
$\mathbf{M}$ and $\mathbf{S}$ are the solutions to the Riccati equations shown in \cite{lqr},
$\log_{2}\left|\det\mathbf{A}\right|$ is the intrinsic entropy rate.
A larger amount of effective information can reduce the achievable lower bound of the LQR cost. 
Therefore, CNER serves as a bridge between heterogeneous SC$^3$-stage capabilities and task-level control performance.

\subsection{Task-oriented systematic optimization}
From the above analysis, it is evident that sensing, communication, computing, and control are tightly coupled, making a systematic joint optimization scheme indispensable.
Traditional D2C design usually optimizes individual links using metrics such as SNR, spectral efficiency, or latency. 
However, such link-level optimization does not fully capture autonomous operations, where sensing, communication, computing, and control jointly determine the task outcome~\cite {link}.

Motivated by the limitations of the traditional link-level paradigm, we adopt a task-oriented perspective for D2C communications between satellites and robots.
Rather than treating the system as a series of isolated stages, we model it as an integrated feedback loop in which all stages continuously interact with each other to update the system state.
The capability disadvantage of one stage can be partially compensated for by the capability enhancement of other stages~\cite{SC3}.
From a task-oriented perspective, CNER can be used as an information-level objective or constraint for system design.
By jointly designing the sensing, communication, computing, and control processes, we can improve the system's task completion performance. 
For example, we can consider sensor capabilities, channel conditions, and computing server capacities, and optimize the allocation of communication and computing resources, data offloading strategies, and time scheduling to achieve the optimal CNER.
In practical operating conditions, this comprehensive, task-oriented optimization scheme enables robots to achieve accurate control and complete operational tasks with maximum efficiency.

\section{Decomposition structures for closed-loop operations}

The D2C satellite communication system based on SC$^3$ closed loops involves information acquisition, transmission, processing, and control stages, resulting in a highly complex system. 
Directly optimizing all sensors, satellites, computing units, and robots may lead to excessive signaling overhead and computational complexity. 
To improve tractability, we decompose the system into finer-grained functional structures according to the dominant bottleneck of the closed loop.

\begin{figure}[ht!]
\centering
\includegraphics[width=3.5in]{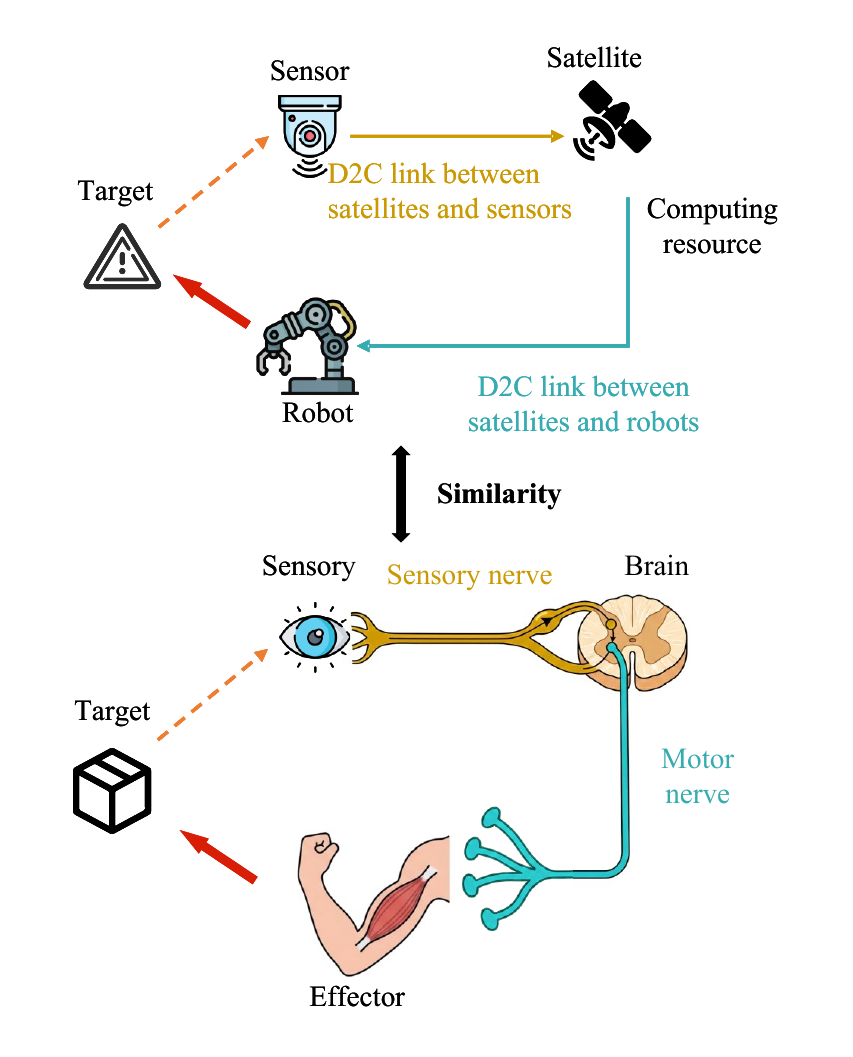}
\caption{Comparison of the reflex arc and the SC$^3$ closed loop.}
\label{fig_arc}
\end{figure}

As shown in Fig.~\ref{fig_arc}, the SC$^3$ closed loop has a functional similarity to the reflex arc.
Sensors correspond to receptors, uplink transmission corresponds to sensory nerves, the computing unit corresponds to the brain, downlink transmission corresponds to motor nerves, and robots correspond to effectors.
It should be noted that the analogy is used only as an intuitive inspiration to highlight feedback, coupling, and functional integrity, rather than implying biological equivalence.
Similar to the reflex arc that is the structural basis for the realization of reflex activities, a single SC$^3$ closed loop is the basic structure of the system for task execution~\cite{arc}.
A single SC$^3$ closed loop is the simplest structure that can continuously learn the behavior of the physical system and steer its evolution toward the desired objectives.
An autonomous task should not be optimized by isolating one link or one device, because the final task performance depends on the coordinated operation of all stages.
However, single closed-loop operations have limitations in sensing, computing, and control stages.
\begin{table*}[!t]
\caption{Comparison of motivations and key issues of three collaborative structures.}
    \begin{tabularx}{\textwidth}{lXXX} 
    \hline
      \textbf{} & \textbf{Motivation} & \textbf{Key issues}\\
      \hline
      \textbf{Collaborative sensing structure}&  Large-area sensing and multi-dimensional state observation.  & Sensor-resource matching; multi-dimensional data fusion.\\
      \textbf{Collaborative computing structure}  &  Limited onboard resources for massive data and complex tasks.  &  LEO/MEO/GEO offloading; resource allocation under channel, ISL, and computing constraints. \\
      \textbf{Collaborative control structure}  &  Multi-robot cooperation for global task objectives.  &  Command generation; robot synchronization; uncertainty and failure handling.\\
    \hline
    \end{tabularx}
    \label{tab:table2}
\end{table*}

\begin{figure*}[ht]
\centering
\includegraphics[width=7in]{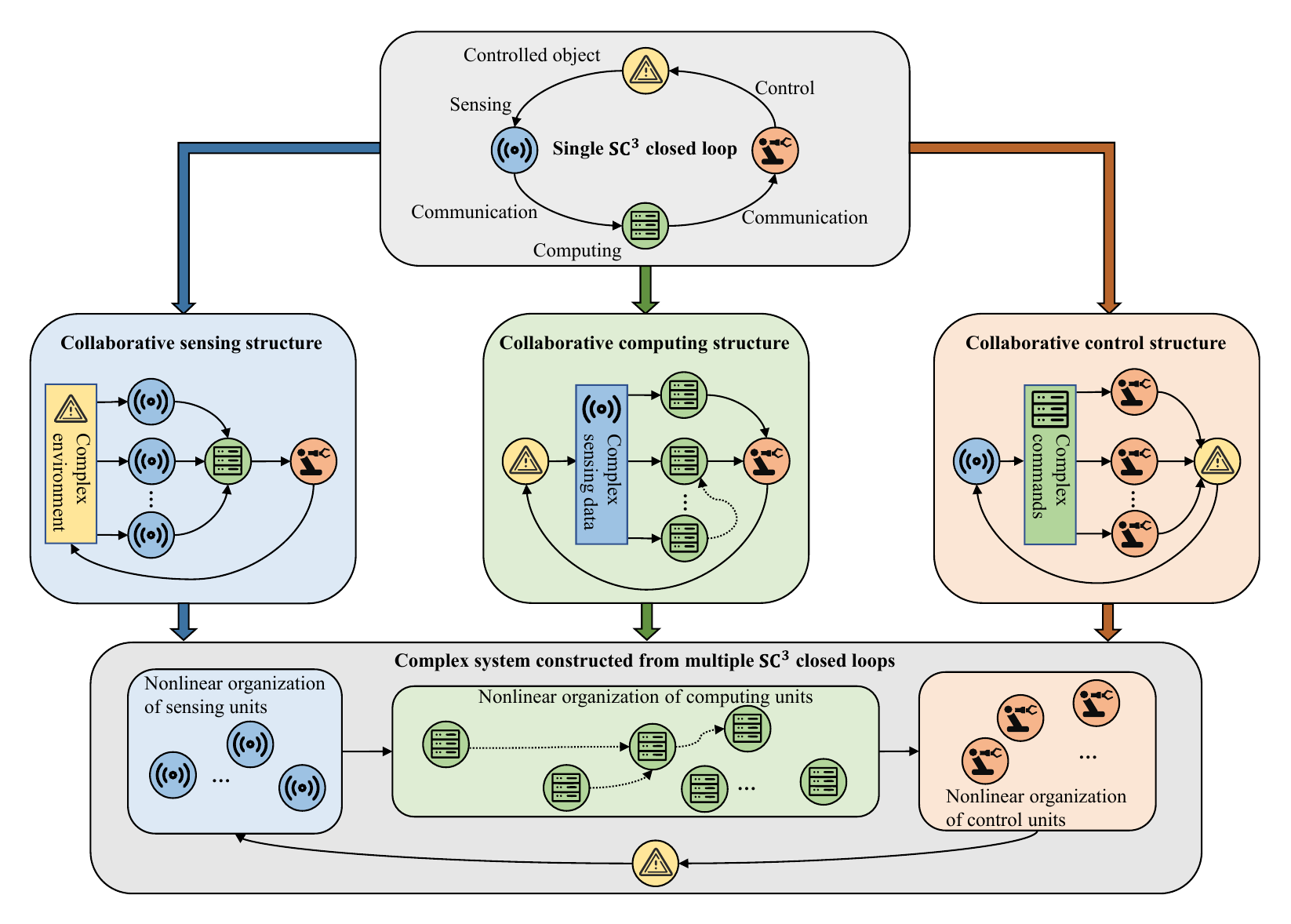}
\caption{Illustration of the single SC$^3$ closed loop, three extended collaborative structures, and the complex system constructed from these structures.}
\label{fig_2}
\end{figure*}

Based on this observation, Fig.~\ref{fig_2} extends the single SC$^3$ closed loop into three representative collaborative structures.
Table~\ref{tab:table2} summarizes their motivations and corresponding key issues. 
When the bottleneck lies in information acquisition, multiple sensors form a collaborative sensing structure.
When the bottleneck lies in data processing, multiple satellites, gateways, or cloud servers form a collaborative computing structure. 
When the bottleneck lies in physical execution, multiple robots form a collaborative control structure.
This decomposition helps transform a large-scale closed-loop optimization problem into several more tractable subproblems, enabling us to better design the D2C satellite communication system to support closed-loop operations.

\subsection{Collaborative sensing structure}
In practical autonomous operation scenarios, precise perception of the operating environment typically requires multiple sensors.
On the one hand, from the perspective of task scale, multiple sensors are needed to collect sufficient information to enable large-area environmental sensing.
On the other hand, from the perspective of task content, different types of sensors are often required to observe multiple indicators to obtain multi-dimensional state information of the controlled objects. 
Thus, a collaborative sensing structure, where multiple sensing units are connected to a common communication, computing, and control unit, is essential in the SC$^3$ closed loop.

In the collaborative sensing structure, key issues can also be considered from the perspectives of task scale and task content.
When sensors cooperate to expand the spatial scope of sensing, effective matching of sensing and communication resources becomes important. 
Due to different channel conditions and sensing capabilities, different sensors collect varying amounts of useful information.
By optimizing sensing data upload strategies that account for these differences, the system’s control performance can be optimized~\cite{sensing}.
Also, extracting valid information from large volumes of sensing data while eliminating redundant or low-quality information improves communication efficiency, enabling the most valuable information to be transmitted using minimal communication resources.
When sensors capture different types of information, the fusion of such multi-dimensional sensing data becomes a key issue. 
By accounting for the intrinsic coupling between data fusion algorithms and computing resource requirements, sensing and computing resources can be jointly allocated, thereby improving the overall system performance.
From the optimization perspective, collaborative sensing converts “uploading all sensing data” into “uploading the most task-relevant information”.
In satellite-assisted systems, different sensors may observe the same target from different locations and experience different D2C channels. 
Therefore, the scheduler can jointly consider sensing quality, channel state, and marginal contribution to CNER when selecting sensors and allocating uplink and computing resources. 
This reduces redundant sensing traffic and improves closed-loop performance under limited satellite resources.

\subsection{Collaborative computing structure}
To handle massive volumes of data and achieve complex task objectives, autonomous operations require adequate computing resources. 
However, the computing capacity available on a single satellite is limited.
When the resources of an individual satellite are insufficient, collaborative computing across multiple satellites becomes necessary.

When using D2C satellite communications for closed-loop operations, collaborative computing relies on task partitioning and data offloading among LEO, MEO, and GEO satellites.
Different satellites play different roles according to their latency, coverage, and resource characteristics. 
LEO satellites are more suitable for latency-sensitive processing because of their shorter propagation delay, but they suffer from limited onboard resources and frequent handovers. 
MEO and GEO platforms can provide broader coverage and more persistent connectivity, but their longer propagation delays make them less suitable for time-critical control loops. 
Therefore, they are more appropriate for regional or global information aggregation, non-real-time task planning, or forwarding data to cloud servers with stronger computing capabilities.

Collaborative computing is also constrained by ISLs. 
Although ISLs enable task migration and cooperative processing among satellites, their capacity, latency, and routing availability vary with satellite topology. 
Consequently, offloading decisions should not only consider computing capability, but also the time needed for ISL transmission, the deadline of the closed-loop cycle, and the size of the generated control commands. 
In this sense, the collaborative computing structure helps to decompose the original closed-loop problem into data partitioning, offloading decision, ISL routing, and resource allocation subproblems, thus providing elastic computing support for dynamic and evolving task demands.

\subsection{Collaborative control structure}
In complex operational scenarios, the capability of a single robot is often insufficient to accomplish assigned tasks independently. 
Thus, collaborative operation of multiple robots is required. 
In this structure, sensing and computing units are shared to connect to multiple control units, which gives rise to a range of new challenges.

In particular, the tight coupling among robots introduces significant complexity in coordination and control. 
Task-oriented collaboration demands effective allocation of control authority across robots. 
When the computing server is processing data and generating instructions, it is necessary to consider the operational capabilities of each robot. 
This includes how individual robots contribute to the global task objectives while satisfying the constraints such as energy consumption, physical limits, and safety requirements. 
Based on these considerations, appropriate commands are formulated and transmitted to the robots, enabling coordinated actions on the controlled objects.

Moreover, shared sensing and computing units must simultaneously support multiple robots, which substantially increases synchronization requirements.
LEO handover and link variations may cause inconsistent command reception, potentially leading to inconsistent state information and degraded collaborative performance.
The joint optimization of communication and control processes is essential to ensure effective cooperation among multiple robots.

Finally, robustness and adaptability are critical concerns in collaborative operations. 
Environmental uncertainties and potential robot failures necessitate control strategies that can dynamically adapt without compromising system stability. 
Through robot cooperation, failures can be mitigated through mutual compensation, thereby enhancing the robustness and reliability of the closed-loop system.

\section{A case study}
In this section, we will provide a case study of a task-oriented design for the D2C satellite communication system for autonomous operations, intending to illustrate how the task-oriented schemes differ from conventional communication-oriented schemes in SC$^3$ closed loops.

The case study considers a satellite-robot system for task execution, consisting of an LEO satellite with computing resources, multiple sensors, robots, and controlled objects.
Sensors upload target information to the satellite via D2C links, after which the satellite processes the data and delivers control commands to the robots.
The actuating robots subsequently execute actions in accordance with the received commands.

For the $k$-th closed loop, we adopt a discrete-time linear control model,
\begin{equation}
    \mathrm{\textbf{x}}_{k,t+1}=\mathrm{\textbf{A}}_k \mathrm{\textbf{x}}_{k,t}+\mathrm{\textbf{B}}_k \mathrm{\textbf{u}}_{k,t}+\mathrm{\textbf{v}}_{k,t},
\end{equation}
where $\mathrm{\textbf{A}}_k$ is the state transition matrix, $\mathrm{\textbf{B}}_k$ is the control input matrix,
$\mathrm{\textbf{x}}_{k,t}$ is the system state, $\mathrm{\textbf{u}}_{k,t}$ is the control input, and $\mathrm{\textbf{v}}_{k,t}$ is the noise at time $t$.
The task's control performance is evaluated by the LQR cost, 
\begin{equation}
    l_k= \sup\lim_{N\to\infty}\mathbb{E}\left[\frac{1}{N}\sum_{t=1}^{N}(\mathrm{\textbf{x}}_{k,t}^{T} \mathrm{\textbf{Q}}_k \mathrm{\textbf{x}}_{k,t}+\mathrm{\textbf{u}}_{k,t}^{T} \mathrm{\textbf{R}}_k \mathrm{\textbf{u}}_{k,t})\right],
\end{equation}
where $\mathrm{\textbf{Q}}_k$ and $\mathrm{\textbf{R}}_k$ are weighting matrices for state deviation and control energy, respectively.

The simulation parameters of this case study are listed as follows~\cite{CNER},~\cite{factor}. 
The LEO satellite's antenna gain is $38.5$ dBi, while the robotic terminal's is $0$ dBi.
The uplink and downlink powers are set to $0.2$ and $20$ W, respectively. 
Ka-band is adopted in the D2C system, with the uplink carrier frequency of $30$ GHz and downlink carrier frequency of $20$ GHz. 
The height of the LEO satellite is assumed to be $600$ km. 
The elevation angles of $5$ robots are randomly generated from $30^\circ$ to $90^\circ$, and the satellite geometry is assumed quasi-static within one closed-loop cycle.
Control-related parameter $\mathrm{\textbf{Q}}_k$ is set to $\mathrm{\textbf{I}}_{100}$, and $\mathrm{\textbf{R}}_k$ is set to $\mathrm{\textbf{0}}$.
The intrinsic entropy rate is set to $50$.
The computing capacity of the LEO satellite is set to $20$ GC/s, and processing one bit of data requires 100 CPU cycles. 
The information extraction ratio is assumed to be $0.1\%$, indicating that 1 bit of task-relevant information is extracted from 1000 bits of sensing data. 
The total time allocated to communication and computing in one closed-loop cycle is $0.2$ s, including transmission, propagation, and computing latency.

For the considered SC$^3$ closed loop, the minimum combined uplink and downlink propagation delays are approximately 4 ms for the considered LEO satellite at 600 km, 66.7 ms for a representative MEO satellite at 10,000 km, and 238.6 ms for a GEO satellite at 35,786 km. 
Given its shorter propagation delay, we select a LEO satellite for the case study to investigate autonomous operations requiring rapid feedback. 
MEO and GEO satellites may support tasks with more relaxed response requirements, such as supervisory control and long-term planning.

\begin{figure*}[!t]
\centering
\subfloat[Single-loop LQR costs under different uplink and downlink band-\\width allocation schemes.]{
\includegraphics[width=3.5in]{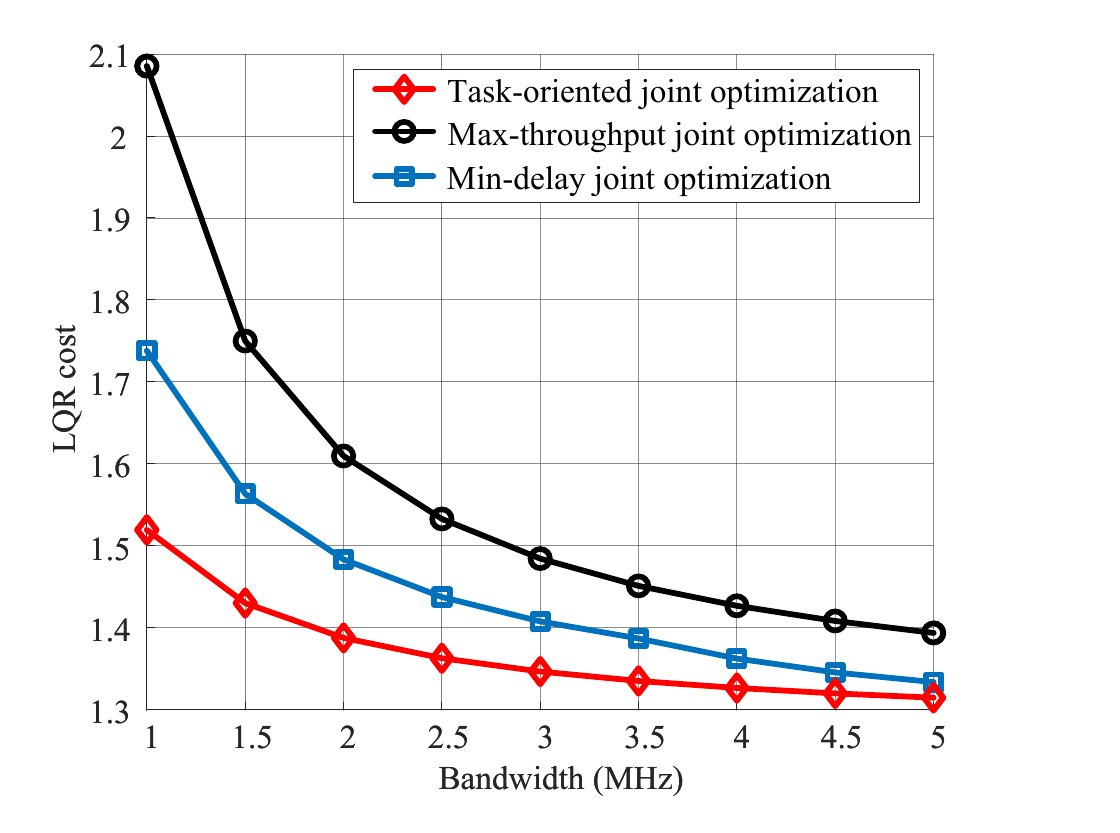}%
\label{fig_4}}
\hfil
\subfloat[Single-loop LQR costs under different information extraction \\ratios achieved with different schemes]{\includegraphics[width=3.5in]{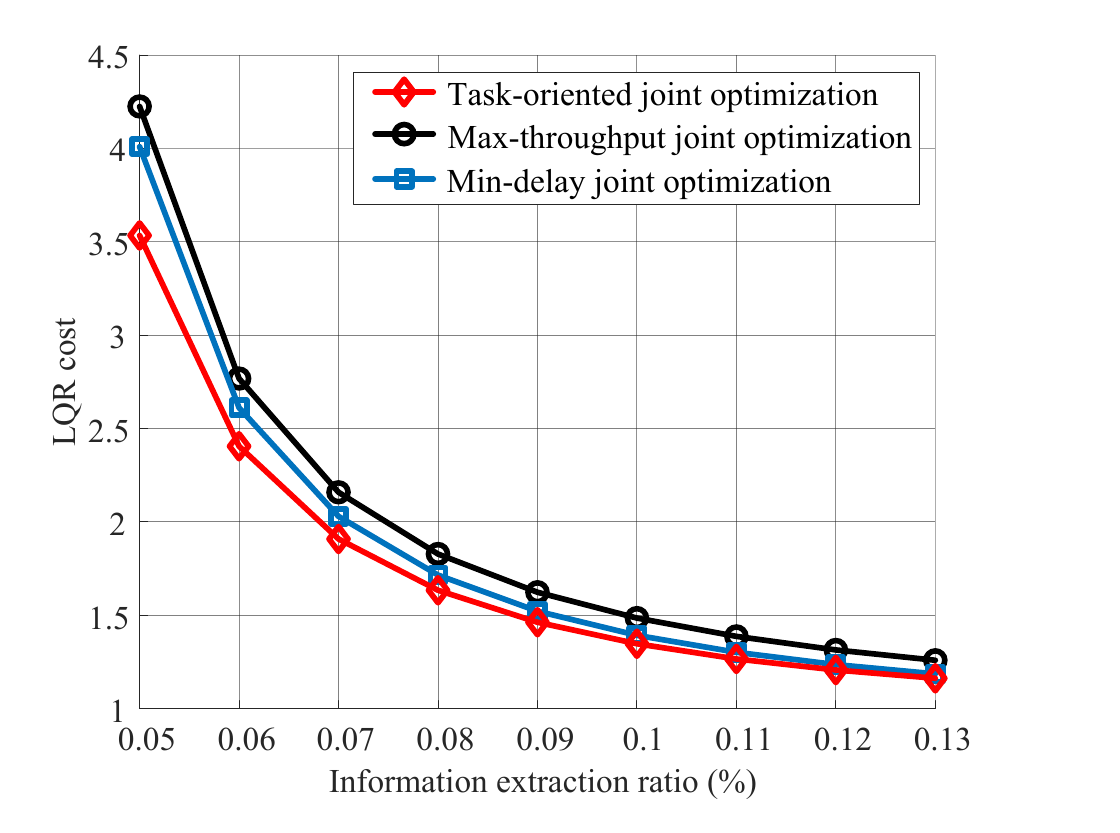}%
\label{fig_5}}
\hfil
\subfloat[Multi-loop LQR costs under different transmit powers achieved \\with different schemes. Error bars indicate 95\% confidence intervals.]{\includegraphics[width=3.5in]{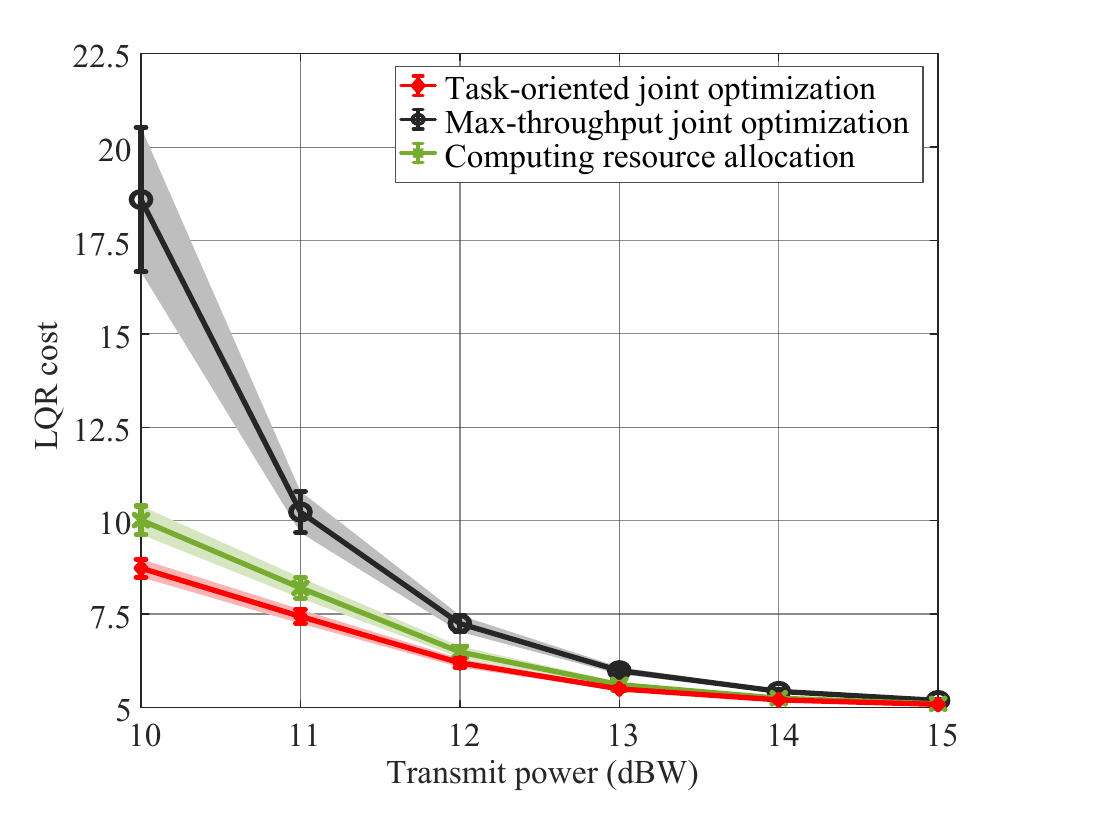}%
\label{fig_6}}
\hfil
\subfloat[Multi-loop LQR costs under different elevation angles achieved \\with task-oriented and max-throughput schemes.]{\includegraphics[width=3.5in]{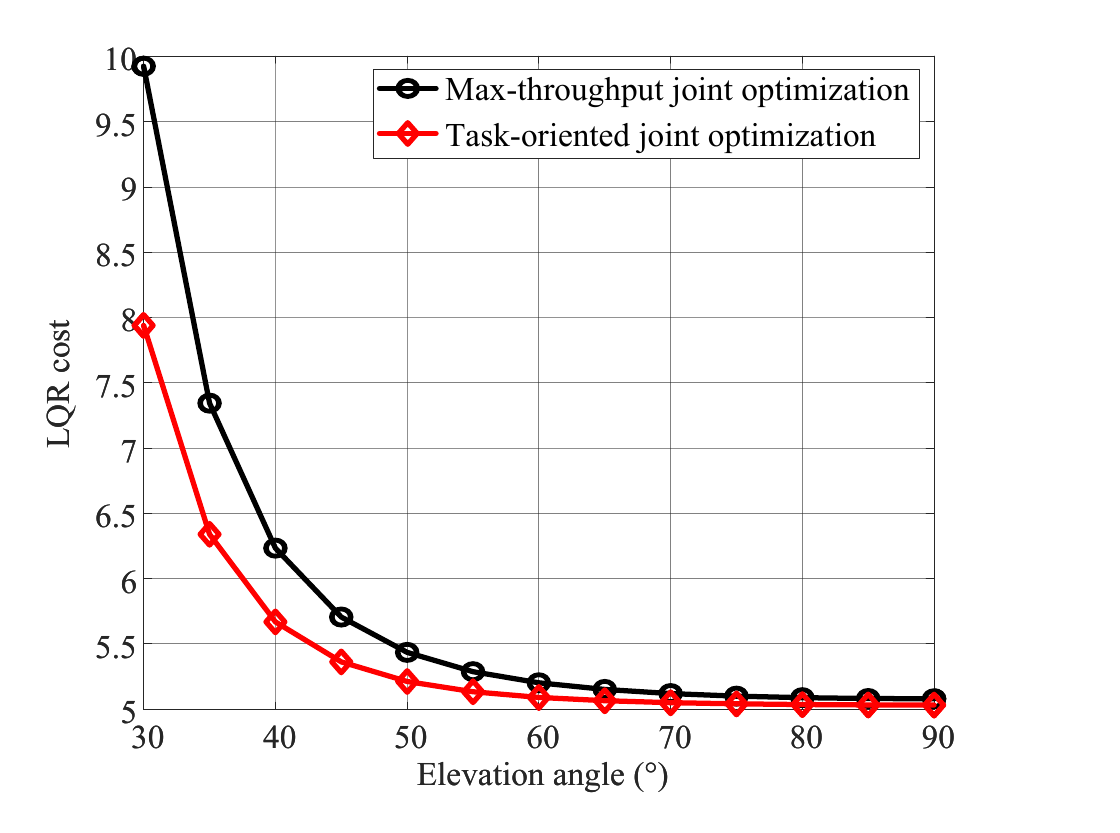}%
\label{fig_7}}
\caption{Simulations of the case study.}
\label{fig_sim}
\end{figure*}

First, we consider the communication resource allocation in a basic single SC$^3$ closed loop. 
Referring to~\cite{CNER}, we analyze the system’s uplink and downlink bandwidth allocation.
We compare the proposed task-oriented scheme with traditional link-level schemes. 
In the task-oriented scheme, the optimization objective is the LQR cost, which is related to the CNER.
In the max-throughput scheme, we take the sum of the uplink and downlink rates as the objective~\cite{CNER}. 
In the min-latency scheme, the objective is to minimize the sum of the uplink and downlink transmission latency.
As shown in Fig.~\ref{fig_4}, the task-oriented scheme achieves the lowest LQR cost, indicating the best control performance, while the min-latency scheme serves as a suboptimal alternative.
In contrast, the max-throughput scheme exhibits the poorest control performance.
The simulation result demonstrates that the proposed task-oriented optimization scheme provides performance advantages, effectively compensating for the limitations of traditional communication-oriented optimization approaches that overlook task completion performance.
Fig.~\ref{fig_5} shows the impact of the information extraction ratio $\rho$.
A larger $\rho$ improves the closed-loop performance by providing more task-relevant information for control decisions.
Moreover, the proposed task-oriented optimization achieves a lower LQR cost than communication-oriented schemes, especially when the information extraction ratio is limited. 
This is because when sufficient information can be extracted, the closed-loop performance becomes less constrained by information availability.

In addition, we consider the joint optimization of communication and computing resources across multiple closed loops.
Referring to~\cite{vtc}, we optimize the allocation of downlink transmit power and computing frequency.
We compare the proposed task-oriented joint optimization scheme with the max-throughput joint optimization scheme and the exclusive computing resource optimization scheme under equal communication resource allocation.
Each point is averaged over 200 independent Monte Carlo trials, where the channel conditions are randomly generated in each trial.  
In Fig.~\ref{fig_6}, error bars indicate 95\% confidence intervals for the mean.
As illustrated in Fig.~\ref{fig_6}, the proposed task-oriented joint optimization scheme achieves a lower LQR cost than the baselines, especially in low transmit-power regimes. 
This result indicates that, when communication resources are limited, allocating them according to control requirements is more effective than purely maximizing communication throughput or minimizing latency.
In Fig.~\ref{fig_7}, we show the impact of the satellite's elevation angle variation on the LQR cost.
The results show that a smaller elevation angle leads to a longer propagation distance and more severe path loss, thereby increasing the difficulty of delivering effective control information.
The task-oriented scheme consistently achieves a lower LQR cost than the max-throughput scheme.
This suggests that the task-oriented design is more robust to unfavorable satellite-terminal environments.
It should be noted that this case study is an illustrative evaluation under simplified assumptions rather than a large-scale system simulation, where detailed beam management, satellite handover and full constellation-level scheduling are not explicitly modeled and remain important directions for future work.

\section{Open research issues}
Although adopting a task-oriented, entropy-based optimization framework and decomposing complex systems can substantially enhance the performance of D2C satellite communications for closed-loop operation, several challenges remain to be addressed.                             

\subsection{Security issues}
Security is particularly critical for D2C satellite communications in closed-loop autonomous operations, because attacks may directly affect task execution. 
Unlike conventional communication services, an SC$^3$ closed loop contains tightly coupled sensing, communication, computing, and control stages. 
As a result, an attack on one stage may propagate through the loop and finally lead to unsafe robot actions or degraded control stability.
In the uplink communication stage, eavesdropping, interference, or data tampering may reduce the reliability of sensing information.
In the downlink stage, command modification may directly misguide robots.
Even without changing the data content, delay attacks can be harmful, since delayed sensing data or control commands may not satisfy the time requirements of the closed loop and increase the control cost.
Thus, security considerations are critical in D2C satellite communications to ensure the reliable and accurate delivery of information to robots.
Leveraging the unique characteristics of satellite communication channels for data encryption and enhancing physical-layer security mechanisms may help mitigate the security risks.

\subsection{Handling system dynamics}
In practical D2C satellite communication systems, networks operate under highly dynamic conditions, where satellite topology, channel conditions, and available resources vary continuously over time. 
During satellite handover, task states, computing context, buffered sensing data, and robot status should be transferred in time.
Otherwise, the new serving satellite may generate outdated or inconsistent control commands.
Future task-oriented designs should jointly consider channel uncertainty, handover probability, onboard resource limits, ISL variations, and closed-loop requirements.
Effectively modeling, predicting, and managing the coupled dynamics arising from system variability is essential for improving the system performance and maintaining overall system stability.

\subsection{Interplay with digital twins}
As a virtual replica of the physical network, a digital twin can monitor and predict channel conditions, computing loads, and control performance. 
Since satellite motion follows predictable orbital dynamics, digital twins may support task-oriented resource management under network variation. 
Key challenges include timely state synchronization with limited communication resources, modeling the coupling among SC$^3$ stages, and maintaining robustness to environmental uncertainty. 
Adaptive digital twins that co-evolve with the physical closed loop are therefore a promising direction for large-scale autonomous operations.

\section{conclusion}
In this work, we have investigated D2C satellite communications for 6G closed-loop operations. 
The system was modeled as SC$^3$ closed loops and characterized from an entropy perspective. 
To manage the system's complexity, we decomposed the overall system into collaborative sensing, computing, and control structures. 
By investigating the core problems within each structure, we gained a clearer and more systematic understanding of D2C satellite communications for 6G closed-loop operations. 
A case study was presented to demonstrate the potential control-performance gain of the task-oriented design over traditional communication-oriented design. 
Finally, we discussed several open research challenges that must be addressed to enable practical industrial deployment.

\section{Acknowledgment}
This work was supported in part by the National Natural Science Foundation of China under Grant 62425110, Grant 62541104, Grant 62571125, and Grant U22A2002, in part by the Key Technologies R\&D Program of Jiangsu (Prospective and Key Technologies for Industry) under Grants BE2023022 and BE2023022-1, and in part by the FAW Jiefang Automotive Co., Ltd.

\bibliographystyle{IEEEtran}
\bibliography{ref}

\section{Biographies}


\begin{IEEEbiographynophoto}{Daohong Shen} (sdh23@mails.tsinghua.edu.cn)
	received the B.S. degree from the Department of Electronic Engineering, Tsinghua University, Beijing, China, in 2023, where she is currently pursuing the Ph.D. degree. Her research interests include space-air-ground integrated networks, task-oriented communications, and 6G mobile communications.
\end{IEEEbiographynophoto}

\begin{IEEEbiographynophoto}{Wei Feng} (Senior Member, IEEE) (fengwei@tsinghua.edu.cn)
	received the B.S. and Ph.D. degrees from the Department of Electronic Engineering, Tsinghua University, Beijing, China, in 2005 and 2010, respectively. He is currently a Professor with 
	the Department of Electronic Engineering, Tsinghua University. His research interests include 6G
	mobile communications, space-air-ground integrated networks, and Internet of Embodied Al Agents.
\end{IEEEbiographynophoto}

\begin{IEEEbiographynophoto}{Yunfei Chen} (Fellow, IEEE) (yunfei.chen@durham.ac.uk)
	received the B.E. and M.E. degrees in electronics engineering from Shanghai Jiaotong University, Shanghai, China, in 1998 and 2001, respectively, and the Ph.D. degree from the University of Alberta in 2006. He is currently a Professor with the Department of Engineering, University of Durham, U.K. His research interests include wireless 
	communications, performance analysis, and joint radar communications designs.
\end{IEEEbiographynophoto}

\begin{IEEEbiographynophoto}{Yongxu Zhu} (Senior Member, IEEE) (yongxu.zhu@seu.edu.cn)
    received the Ph.D. degree in electrical engineering from University College London, in 2017. From 2017 to 2019, she was a Research Associate at Loughborough University. From 2019 to 2023, she held positions as a Senior Lecturer at London South Bank University and an Assistant Professor at the University of Warwick. She is currently a Professor at Southeast University. Her research interests include wireless network topology, collective intelligence communication networks, cell-free wireless networks, and non-terrestrial networks.
\end{IEEEbiographynophoto}

\begin{IEEEbiographynophoto}{Jinxia Cheng} (chengjinxia@chinamobile.com)
    received the Ph.D. degree in information and communication systems from Tsinghua University, China, in 2006. She currently works for the Department of Satellite-Ground Convergence Technology Research at the China Mobile Research Institute, focusing on the research of 5G network technologies and solutions for vertical markets and the integrated satellite-terrestrial networks.
\end{IEEEbiographynophoto}

\begin{IEEEbiographynophoto}{Dapeng Wang} (wangdapeng@chinamobile.com)
    received the B.S. degree in Electronic Engineering from Xi'an Jiaotong University, Xi'an, China, in 1997, and the M.S. degree from Beijing University of Posts and Telecommunications, Beijing, China, in 2000. He is currently a Group-Level Chief Expert with the China Mobile Research Institute. His research interests include radio frequency (RF) technology research of 5G/6G base stations, as well as the formulation of RF standards, product development, satellite communication innovative solution design, spectrum strategy and system compatibility research.
\end{IEEEbiographynophoto}

\begin{IEEEbiographynophoto}{Shi Jin} (Fellow, IEEE) (jinshi@seu.edu.cn)
    received the Ph.D. degree in information and communications engineering from the Southeast University, Nanjing, in 2007. From 2007 to 2009, he was a Research Fellow with Adastral Park Research Campus, University College London, London, U.K. He is currently with the Faculty of the National Mobile Communications Research Laboratory, Southeast University. His research interests include space-time wireless communications, random matrix theory, and information theory.
\end{IEEEbiographynophoto}

\vfill

\end{document}